\documentclass[letterpaper]{JHEP} 

\usepackage{graphicx}
\usepackage{psfrag}

\providecommand{\qqbar }{\ensuremath{ q\overline{q} }}
\providecommand{\qbar}{\ensuremath{\overline{q}}}
\providecommand{\ibar}{\ensuremath{\overline{\imath}}}

\makeatletter 
\preprint{hep-ph/0110257\\
\@date
}
\makeatother

\title{ NLO corrections in MC event generator for angular distribution of 
         Drell-Yan lepton pair production}

\author{Yujun Chen$^1$\footnote{Now in Physics department, University of 
        California at Davis}
         \hspace{10pt} 
        John Collins$^2$\hspace{10pt} 
        Xiaomin Zu$^3$ 
        \\
        Physics Department,
        Penn State University, \\
        104 Davey Laboratory,
        University Park PA 16802,
        U.S.A. 
\\ \email{$^1$ ychen@lifshitz.ucdavis.edu}
\\ \email{$^2$ collins@phys.psu.edu}
\\ \email{$^3$ xiaozu@phys.psu.edu}
}

\date{12 October 2001}


\abstract{
        Using a subtraction method, we derive the formulae suitable for 
use in  Monte-Carlo event generators to give 
the angular distribution for 
the gluon-quark induced NLO corrections in Drell-Yan lepton pair production. 
We also give the corresponding helicity density matrix for $W$ and $Z$ boson
production.

}

\begin{document}

\section{Introduction}
\label{sec:intro}

In Ref.\ \cite{YJ} we applied the subtractive method of \cite{JC,JC1} to
derive the NLO gluon-quark-induced contribution for the Drell-Yan
process in a form suitable for use in a Monte-Carlo event generator.  In this
paper we extend these results to provide 
the angular distribution for the leptons, and,
equivalently, the helicity density matrix for $W$ and $Z$ boson
production. 

As described in \cite{YJ,JC,JC1}, an event 
generator using this subtraction method generates two classes 
of events: One class is 
obtained from the leading order (LO) 
subprocess by showering the initial and final state quarks; the other class 
is generated by starting with a subtracted next leading order (NLO) 
subprocess. As in \cite{YJ}, we only treat 
the gluon-quark induced subprocesses, 
$qg  \to Vq' \to q'll'$  ($V= \gamma^*,\, W^\pm$ or $Z^0$). 
This NLO subprocess is not necessarily suppressed by a factor 
of $\alpha_s$ if  the gluon distribution function is larger than 
the quark distribution function.
The quark-antiquark induced NLO corrections are complicated by the 
presence of soft divergences \cite{JC2}, so we defer their detailed 
treatment in our method to the future. 

The polarization state of the gauge boson ($\gamma^*$, $W^\pm$ or $Z^0$)
determines the angular distribution of the decay leptons in the
lepton pair  rest frame. The general structure of the angular distribution
is given by nine  helicity cross sections corresponding to the nine 
helicity density matrix elements for the gauge boson \cite{Mirkes}. 
In the LO parton-level subprocesses ($\qqbar \to V \to ll'$) 
only the transverse polarization of
gauge boson contributes to the helicity density matrix. In NLO, 
in general, all three polarizations of the gauge boson contribute, but 
the subtraction term, associated with a collinear divergence, only
involves the same transverse polarization as in LO. For the 
$\gamma^*$, only the 4 parity-conserving terms in the cross 
section  are 
nonzero \cite{LamTung}; but for $W^\pm$ and $Z^0$, with their
characteristic $V-A$ and $V+A$ couplings, there are two other
nonzero helicity cross sections \cite{Mirkes}.

The organization of this paper is the following. 
In Sec.\ \ref{sec:algorithm}, we describe the treatment of lepton pair
production in an event generator.  Then we compute the effect of combining the
LO parton-level cross section with the order $\alpha_s$ part of the 
shower; this will be needed as the subtraction term in the NLO hard cross 
section calculation. 
In Sec.\ \ref{sec:NLO correction}, we carry out the subtraction from 
NLO helicity cross sections to get the angular distribution in the
NLO differential cross sections.
For $W^\pm$ and $Z^0$ production, we also give the helicity density matrix.
Finally, we summarize and discuss our results in Sec.\ \ref{sec:conclusion}.

\section{Monte-Carlo algorithms}
\label{sec:algorithm}


\subsection{QCD improved parton model and inclusive lepton pair production}

Our calculations are based on the QCD improved parton model, which
shows that a hard scattering process initiated by two hadrons is the
result of an interaction between their constituents, namely  quarks and 
gluons \cite{text}.

We consider the process $A + B \to V + X \to ll' + X$  
in a collision of two hadrons  $A$ and $B$ at a center-of-mass energy-squared
 $s=(P_A+P_B)^2$. When the momentum fractions carried by the partons are 
not too small (roughly $\geq 0.01$), LO subprocesses 
($\qqbar \to V \to ll'$) are dominant, with the NLO subprocesses being
suppress by a factor of $\alpha_s(Q^2)$.  But at the higher energies of the Fermilab 
Tevatron  ($\sqrt{s} = 1.8$ TeV), the momentum fractions are often smaller,
and the gluon density can be larger than the quark and antiquark
densities, so that gluon-induced NLO subprocesses like $qg \to Vq' \to q'll'$
are not necessarily suppressed by $\alpha_s$ relative to the LO subprocess.

 The LO cross section
$d\sigma^{\rm LO}(V)/dQ^2dyd\Omega$ for producing a lepton pair with 
rapidity $y$, invariant mass-squared $Q^2$ and relative solid angle 
 $\Omega=(\theta,\phi)$ in the Collins-Soper frame \cite{CS}
is obtained by weighting the parton level cross section with the parton
distribution functions (pdf's) $f_{i/A}(x_a, Q^2)$ and 
$f_{\ibar/B}(x_b, Q^2)$ \cite{text}:
\begin{eqnarray}
  \label{LO}
      \frac{d\sigma^{\rm LO}(V)}{dQ^2dyd\Omega}
     &=& \sum_{q \qbar'} \hat{\sigma}(V)
         \mathcal{D}_{\qqbar'}(\theta)
        f_{q/A}(x_a, Q^2)f_{\qbar '/B}(x_b, Q^2)
\nonumber \\
        && +  \sum_{\qbar'q} \hat{\sigma}(V)
        \mathcal{D}_{\qbar 'q}(\theta)
        f_{\qbar '/A}(x_a, Q^2)f_{q/B}(x_b, Q^2) ,
\end{eqnarray}
where
\begin{eqnarray}   
         \mathcal{D}_{\qqbar'}(\theta)&=&
         (1+\cos^2 \theta + 2A_qA_l \cos \theta), 
\nonumber \\
         \mathcal{D}_{\qbar 'q}(\theta)&=&\mathcal{D}_{\qqbar'}(\pi-\theta) .
\end{eqnarray}
The parton-level cross section $\hat{\sigma}(V)$ and 
the coefficients $A_q$ and $A_l$ depend on the boson being produced:
\begin{itemize}
\item $V=\gamma^*$:
\begin{equation}   
\label{sigma.gamma}
        \hat{\sigma}(\gamma^*) = \frac{3}{16\pi} \times
        \frac{4\pi \alpha^2 e_q^2}{9Q^2s}, ~ A_q=A_l=0, 
\end{equation}
where $e_q$ is the quark charge in units of the size of the electron
charge $e$. 
The summation is over $(q, \qbar')$ in
$\{ (u, \overline{u}),\, (d, \overline{d}),\, 
(s, \overline{s}),\, (c, \overline{c})\}$, where we restrict our
attention to quarks that can be approximated as massless in the
production of $W$ and $Z$ bosons.

\item $V=W^\pm$:
\begin{eqnarray}
\label{sigma.W}
        \hat{\sigma}(W^\pm) &=& \frac{3}{16\pi}\times
         \frac{4\pi \alpha^2}{9s} 
                \frac{|V_{\qqbar'}|^2}{16s_w^4} \times 
                \frac{Q^2}{(Q^2-M_W^2)^2+M_W^2\Gamma_W^2} 
\nonumber \\
        &\cong& \frac{3}{16\pi} \times
        \frac{\pi}{3} \sqrt{2} G_F  |V_{\qqbar'}|^2 
         \frac{Q^2}{s} \delta(Q^2-M_W^2) B(W \to l\nu)
\\ 
        && W^+: A_q=A_l=1, ~~ W^-: A_q=-A_l=1,
\nonumber
\end{eqnarray}
where $V_{\qqbar'}$ is an CKM matrix element, $s_w = \sin \theta_w$,
$G_F$ is the 
Fermi coupling constant, and $B(W \to l\nu)$ is the branching ratio 
$\Gamma(W \to l\nu) / \Gamma_W$. 
In the second line of this equation, we have used the narrow width
approximation for the decay of $W$ boson.
For the $W^+$, the summation is over the cases
$(q, \qbar')$ in $\{ (u, \overline{d}),\, (c, \overline{s})\}$,
while for the $W^-$ we have
$(q, \qbar')$ in $\{(d, \overline{u}),\, (s, \overline{c})\}$. 

\item $V=Z^0$:
\begin{eqnarray}
\label{sigma.Z}
        \hat{\sigma}(Z^0) &=& \frac{3}{16\pi} \times 
         \frac{4\pi \alpha^2}{9s} 
                \frac{[q_V^2+q_A^2][l_V^2+l_A^2]}{16s_w^4c_w^4} \times 
                \frac{Q^2}{(Q^2-M_Z^2)^2+M_Z^2\Gamma_Z^2} 
\nonumber \\
        &\cong& \frac{3}{16\pi} \times 
        \frac{\pi}{3} \sqrt{2} G_F  [q_V^2+q_A^2]
         \frac{Q^2}{s} \delta(Q^2-M_Z^2) B(Z^0 \to l\overline{l})
\\
        && A_q=\frac{2q_Vq_A}{q_V^2+q_A^2}, ~~~
         A_l = \frac{2l_Vl_A}{l_V^2+l_A^2},
\nonumber
\end{eqnarray}
where $q_A = T_q^3$, $q_V=T_q^3-2e_qs^2_w$, $c_w = \cos \theta_w$, $s_w = \sin \theta_w$,
$B(Z^0 \to l\overline{l})$ is the branching ratio for $Z^0$, and the
summation is the same as in the case of $\gamma^*$.  In the second line,
we have again used the narrow-width approximation.
\end{itemize}
The fractional momenta assigned to the incoming quark and antiquark
are equal to the following kinematic variables:
\begin{equation}
     x_a = e^y \sqrt{\frac{Q^2}{s}}, ~~~
     x_b = e^{-y} \sqrt{\frac{Q^2}{s}}.
\end{equation}

With full perturbative QCD corrections taken into account, Eq.\
(\ref{LO}) is replaced by the factorization formula, 
\begin{equation}
\label{factorization}
     \frac{ d\sigma }{dQ^2 dy d\Omega} 
         = \sum_{i,j} \int d\xi_i \, d\xi_j \, 
    f_{i/A}(\xi_i,\mu^2) f_{j/B}(\xi_j,\mu^2) 
        \frac{ d\hat{\sigma}_{ij}}{dQ^2 dy d\Omega} ,
\end{equation}
where the $\xi$'s are the momentum fractions of the incoming partons,
and $d\hat\sigma_{ij}/dQ^2dyd\Omega$ is a suitably constructed parton-level
hard scattering cross section, which depends on $x_a/\xi_i,\ x_b/\xi_j$,
the vector boson mass $Q$, the solid angle $\Omega$, and on the
renormalization/factorization scale $\mu$ through $\mu/Q$ and
$\alpha_s(\mu)$.  Now the sum is over all pairs of parton flavors (quarks
and gluons).  The formal domain of validity of Eq.\ 
(\ref{factorization}) is the asymptotic `scaling' limit, analogous to
the Bjorken limit in DIS, where $s, Q^2 \to\infty$ with $x_a$ and $x_b$
fixed.

\subsection{Parton-shower algorithm}
\label{sec:shower}
The initial-state shower algorithm for vector boson production used in
PYTHIA or RAPGAP is described in \cite{BSZ} (see also
\cite{YJ,JC}). We use the method described in \cite{YJ} to define the
splitting variable $z$, rather than the $\hat{s}$ method of
\cite{BSZ}, since the treatment of NLO corrections is simpler.

The part of the algorithm used in an event generator that concerns us
is the following:
\begin{enumerate}

\item Generate values of $y$ and $Q^2$, then generate a direction
   (i.e., $\Omega$) for the leptons according to  
   the LO cross section for lepton pair production, Eq.\ (\ref{LO}).
   This gives the initial values of the parton and lepton 4-momenta.
   (These initial values have zero transverse momentum for the partons 
   and hence for the lepton pair.  After showering, the parton
   4-momenta will be modified.)  The direction $\Omega$ gives polar angles 
   $\theta$ and $\phi$ that we choose to interpret in the Collins-Soper
   frame \cite{CS}.

\item For branching on hadron $A$ side:
\label{Q1}
   Generate a virtuality $Q_1^2$ for the incoming quark 
   $i$, the first
   longitudinal momentum splitting variable $z_a$, and
   an azimuthal angle $\phi'$ for this branching. The distributions 
   arise  from the Sudakov form factor
\begin{eqnarray}
\label{Sudakov}
   S_i(x_a, Q_{\rm max}^2, Q_1^2) 
   &=& \exp\left\{
      - \int_{Q_1^2}^{Q_{\rm max}^2} \frac{dQ'^2}{Q'^2}
       \frac{\alpha_s(Q'^2)}{2\pi} \right.
\\          
   & & \hspace{27pt} \times
      \sum_k \int_{x_a}^1 \frac{dz_a}{z_a} \, P_{k\to ij}(z_a) 
                        \frac{f_k(x_a/z_a,Q'^2)}{f_i(x_a,Q'^2)}
      \Bigg \}.
\nonumber
\end{eqnarray}
  Here, $Q_{\rm max}^2$ is normally set equal to 
   $M^2$ for $W^\pm$ and $Z^0$, and for $\gamma^*$ it is often fixed to be
   $Q^2$ --- the hard scattering scale.
   The Sudakov form factor is the probability that the virtuality of
   quark $i$ is less than $Q_1^2$. The branching on hadron $B$ side is done 
   similarly with first splitting variable $z_b$.

\item Iterate the branching for all initial-state and
   final-state\footnote{
      The final-state showering is organized similarly to the
      initial-state showering.  However, we will not need it
      explicitly in this paper.
   }
   partons until no further branchings are possible.

\item Compute the actual values of the parton and lepton 4-momenta.

\end{enumerate}


\subsection{Subtraction term from first initial-state branching}
\label{sec:subtraction}
\FIGURE{
\centering
\psfrag{A}{$A$}
\psfrag{B}{$B$}
\psfrag{q}{$q$}
\psfrag{p_1}{$p_1$}
\psfrag{p_2}{$p_b$}
\psfrag{p_3}{$p_a$}
\psfrag{p_1'}{$p_1'$}
\psfrag{s}{$\hat{s}$}
\psfrag{l}{$p_l$}
\psfrag{l'}{$p_{l'}$}
\includegraphics [scale=0.30]{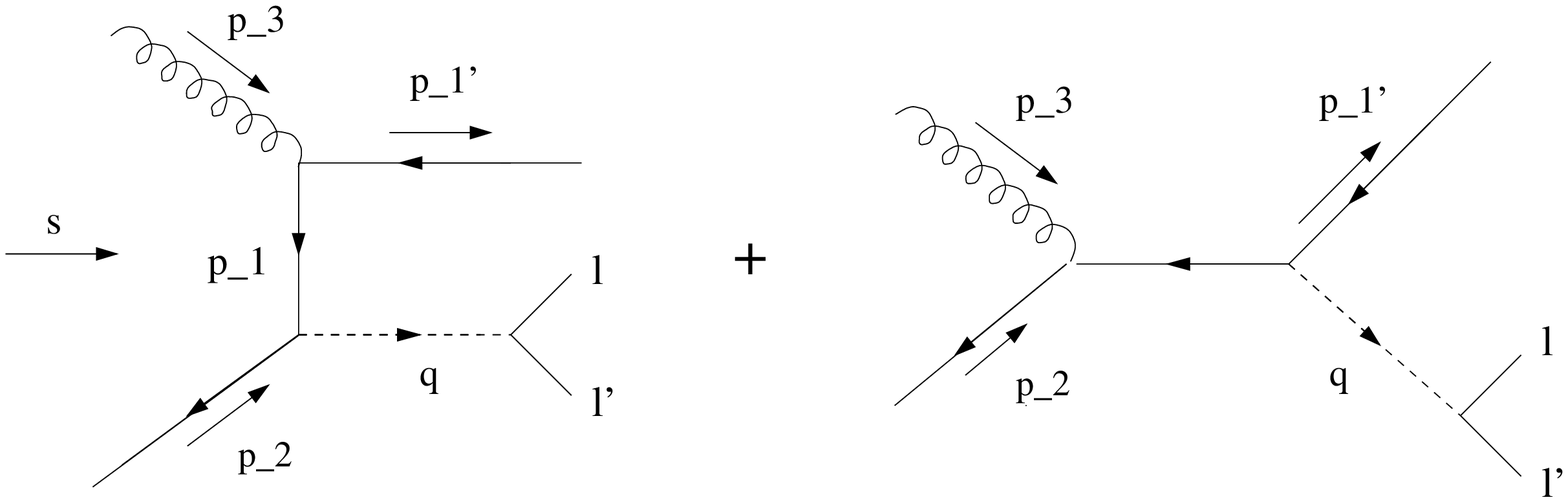}
\caption{NLO gluon-quark collision. Gluon comes from hadron $A$ and
  (anti)-quark comes from hadron $B$.}
\label{fig:gluon-quark.1}
}

In Sec.\ \ref{sec:NLO correction}, we will calculate the
hard scattering cross section for $gq$-induced process (Fig.\ 
\ref{fig:gluon-quark.1}).  The gluon, of momentum $p_a$, comes from
hadron $A$ and the (anti)-quark, of momentum $p_b$, comes from hadron
$B$.  
The NLO contribution we want to calculate is to be accurate when 
the following conditions are satisfied. 1) The incoming gluon and quark 
have virtualities and transverse momenta-squared that are small compared 
with $Q^2$. 2) The intermediate quark, of momentum $p_1$, has a virtuality of 
order $Q^2$. 3) The lepton pair has a transverse momentum of order
$Q$.  
To avoid double counting, it is necessary to subtract the corresponding 
contribution obtained from the showering algorithm applied to the LO parton 
level cross section. It is this subtraction term 
that we calculate in this section. 
After this subtraction, the combination of 
the LO and NLO term will also be accurate when the transverse momentum 
of the lepton pair or the virtuality of the intermediate quark
decreases. 

The subtraction term is obtained by multiplying the LO parton level
cross section, from Eq.\ (\ref{LO}), by the first order term in the
expansion of the Sudakov form factor Eq.\ (\ref{Sudakov}) in powers of
$\alpha_s(Q^2)$. It is made differential in the momenta of the particles
involved, and the gluon-to-quark splitting kernel is selected. To
match with the definition of the NLO hard-scattering, it should be
calculated when the initial-state partons are on-shell and have zero
transverse momentum, and the final-state quark is on-shell.

That is, we subtract the first-order cross section in the showering
approximation:
\begin{eqnarray}
\label{Collinear.gq}
   \frac{d\sigma_{\rm shower\, 1}(V)}
        {dQ^2 \, dy \, d\Omega \,dQ_1^2 \, d \xi_a \,d\phi'}
&=&     \sum_{q, \qbar'}
            \hat{\sigma}(V) \, \mathcal{D}_{\qqbar'}(\theta)
         \frac{\alpha_s(Q^2)}{4\pi^2 Q_1^2} \,    C_1(Q_1^2) 
         P(z_a)\,  \frac{ 1 }{ \xi_a } f_g(\xi_a, Q^2) f_{\qbar'}(x_b,Q^2)
\nonumber\\
 && \hspace*{-1cm}
   + \,\mbox{case: antiquarks (from $g$) come from $A$, quarks from $B$} .
\end{eqnarray}
Note that in the summation in this equation, we still sum over $q,
\qbar'$.  The gluon from $A$ splits into a $\qqbar$ pair, and in the
first term the quark from the gluon interacts with the antiquark from
$B$.  In the second term, on the last line, we have the antiquark
from the gluon interacting with the quark from $B$.  
\FIGURE{
  \centering \psfrag{P_A}{ $\vec{P_B}$} \psfrag{P_B}{ $\vec{P_A}$}
  \psfrag{X}{$x (\vec{q_T})$} \psfrag{Y}{$y$} \psfrag{Z}{$z$}
  \includegraphics [scale=0.3]{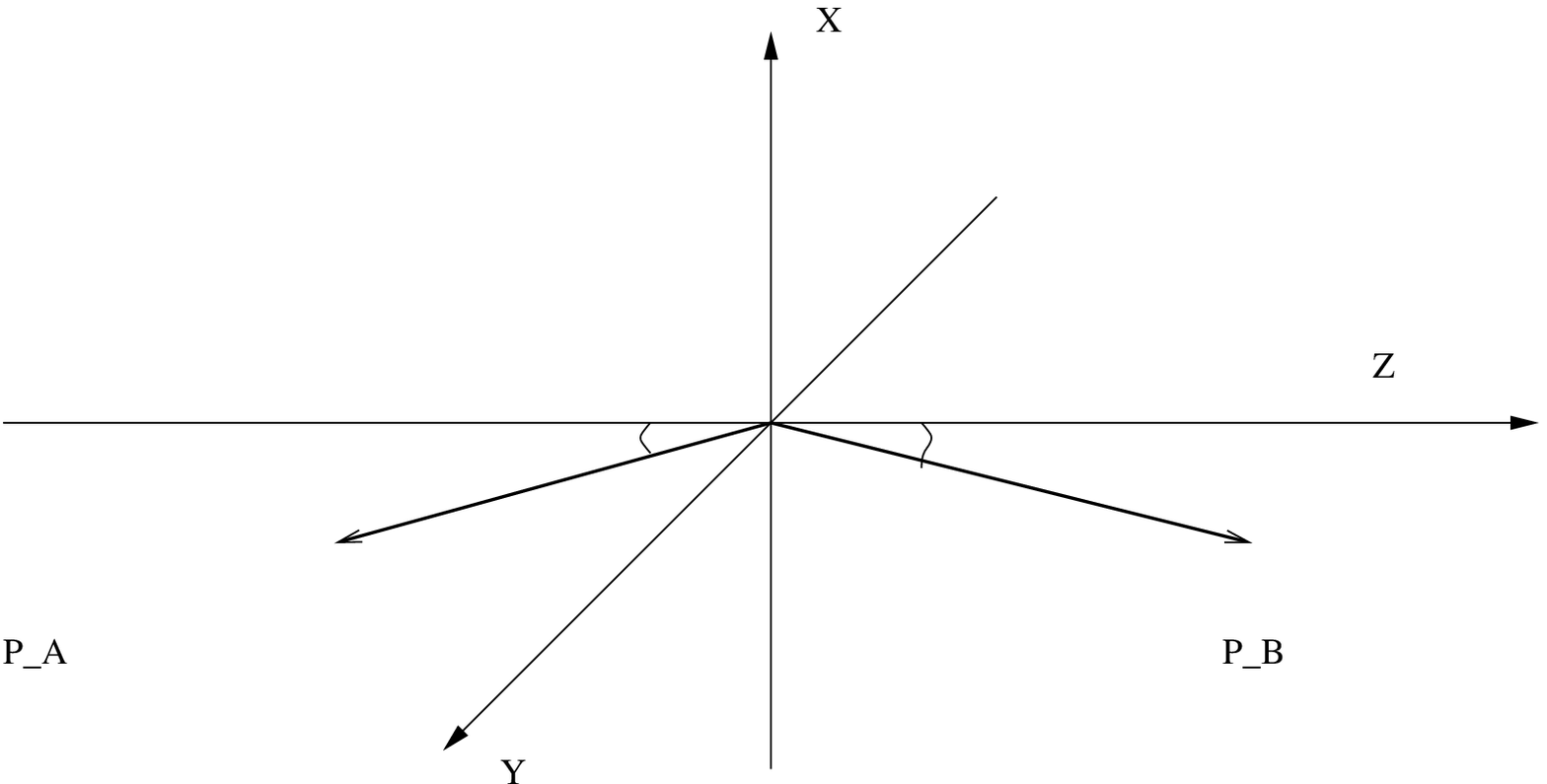}
\caption{Collins-Soper frame. $\vec{P_A}$ and $\vec{P_B}$ lie in the 
    $x$-$z$ plane of the lepton-pair rest frame and z-axis bisects
    $\vec{P_A}$ and $-\vec{P_B}$. }
\label{fig:frame}
}
The splitting kernel here is for $\mbox{gluon}
\to\mbox{quark}+\mbox{antiquark}$: $P(z) = P_{g\to \qqbar}(z) = \frac12
(1-2z+2z^2)$.  Note that because we are doing a strict expansion in
powers of $\alpha_s(Q^2)$, the scale argument of the pdf is $Q^2$.  The
function $C_1(Q_1^2)$ is a cut-off function that gives the maximum
value of $Q_1^2$.  Normally it is $\theta(Q^2-Q_1^2)$, but as discussed in
\cite{JC}, a different choice of the cut-off function could minimize
the number of negative weight events generated generated from the
subtracted NLO cross section.  The general solution
to deal with a smooth cut-off function is given in \cite{JC1} for a
non-gauge theory. We plan to apply this method to QCD in the future.

Now, working in Collins-Soper frame \cite{CS} (see Fig.\ 
\ref{fig:frame}), we reconstruct the 4-vectors for the momenta $q$,
$p_1'$, $p_a$ and $p_b$ of the lepton pair, the outgoing quark, the
incoming gluon and the incoming (anti)-quark. To be consistent with
\cite{YJ}, they obey the following requirements:
\begin{enumerate}
\item The $z$-axis is given by the Collins-Soper \cite{CS} definition
    for the angles of the leptons; it bisects the angle between
    $\vec{P}_A$ and $-\vec{P}_B$, where the momenta are for the incoming 
    {\em hadrons}.  This is shown in Fig.\ \ref{fig:frame}.
    (Bengtsson, Sj{\"o}strand and van Zijl \cite{BSZ} made a different
    choice of axes in the overall CM frame.)

\item   The incoming partons, $p_a$ and $p_b$ have momentum fractions  
    $\xi_a$ and $\xi_b$ relative to their parent hadrons, in the sense
    of light-front components.

\item $p_1^2 = -Q_1^2$.

\item 
\label{ext.partons}
  $p_a^2 = p_b^2 = p_1'^2 = 0$, and $p_a$ and $p_b$ have zero
  transverse momentum in the overall CM frame.

\item $q^2 = Q^2$, where Q is the invariant mass of the lepton pair.

\item $y =  \displaystyle \frac{1}{2} \ln \frac{P_B\cdot q}{P_A\cdot q}
         = \displaystyle \frac{1}{2} \ln \frac {x_a}{x_b}$,
      is the exact rapidity of the lepton pair.

\item The first splitting variable on the $A$ side is defined by
\begin{equation}
\label{za}
        z_a = \displaystyle \frac{x_a}{\xi_a}.
\end{equation}
\end{enumerate}   
Notice that in this definition, the splitting variable $z_a$ is
defined in terms of the kinematic variable 
$x_a = e^yQ/\sqrt{s}$, rather than in terms of the
fractional momentum of the intermediate quark.

We define the Mandelstam variables at the  parton level as usual:
\begin{eqnarray}
\label{s.hat}
         \hat{s}&=&(p_a+p_b)^2=\xi_a\xi_bs
\\
\label{t.hat}
         \hat{t}&=&(p_b-q)^2=
         \xi_a x_a \frac{x_b^2-\xi_b^2}{x_a\xi_b+x_b\xi_a}s = -Q_1^2
\\
\label{u.hat}
         \hat{u}&=&(p_a-q)^2=
         \xi_b x_b \frac{x_a^2-\xi_a^2}{x_a\xi_b+x_b\xi_a}s.
\end{eqnarray}
These satisfy the constraint that $\hat{s}+\hat{t}+\hat{u}-Q^2=0$ .
Then the partons' 4-momenta in Collins-Soper frame are given by:
\begin{eqnarray}
 \label{p.a.mu}
    p_a^\mu &=& \xi_a E_A\, (1,\, \sin \eta,\, 0,\, \cos \eta),
 \\
 \label{p.b.mu}
    p_b^\mu &=& \xi_b E_B\, (1,\, \sin \eta,\, 0,\, -\cos \eta),
 \\
 \label{q.mu}
    q^\mu &=& Q\, (1, \, 0, \, 0, \, 0),
 \\
 \label{p.1'.mu}
    p_1'^{\mu} &=& p_a^\mu + p_b^\mu - q^\mu
 \nonumber
 \\     &=& \left( [\xi_aE_A+\xi_bE_B-Q], \, [\xi_aE_A+\xi_bE_B]\sin \eta,
                 \, 0, \, [\xi_aE_A-\xi_bE_B]\cos \eta
            \right),
 \\
 \label{p.l.mu}
    p_l^\mu &=& \frac{Q}{2}\, (1, \, \sin \theta \cos \phi, \,
                    \sin \theta \sin \phi, \, \cos \theta),
 \\
 \label{p.l'.mu}
    p_{l'}^\mu &=& \frac{Q}{2}\, (1, \, -\sin \theta \cos \phi, \,
                    -\sin \theta \sin \phi, \, -\cos \theta),
 \end{eqnarray}
 with
 \begin{eqnarray}
 \label{Eandr}
         E_A&=&\frac{\sqrt{s}}{2} \sqrt{1+r^2} e^{-y}, ~
         E_B=\frac{\sqrt{s}}{2} \sqrt{1+r^2} e^{+y},
 \\
         \sin \eta &=& \frac{-r}{\sqrt{1+r^2}}, ~~~~~~~~ 
         \cos \eta =\frac{1}{\sqrt{1+r^2}}, 
 \\
         r&=&\sqrt{\frac{Q_T^2}{Q^2}}=\sqrt{\frac{\hat{u}\hat{t}}{\hat{s}Q^2}}.
 \end{eqnarray}

Convenient  variables for the above subtraction term are $Q^2$, $y$, 
$\Omega$, $\xi_a$ and $Q_1^2$.  However, they are
not so convenient for the NLO matrix-element calculations.  So we  
transform the
 cross section in Eq.\ (\ref{Collinear.gq}) in terms of more convenient
 variables for a hard gluon-quark scattering: $Q^2$, $y$, $\Omega$ ,
 $\xi_a$ and $\xi_b$.  Then the cross section after integrating 
over the branching angle $\phi'$ is: 
\begin{eqnarray}
\label{Collinear.gq1.new}
     \frac{ d\sigma_{\rm shower\, 1}(V) }
          { dQ^2\, dy \, d\Omega \, d\xi_a \, d\xi_b }
  &=&     \sum_{q, \, \qbar'} \frac{\alpha_s(Q^2)}{2\pi}\,
            \hat{\sigma}(V) \, \mathcal{D}_{\qqbar'}(\theta)
        C_1(Q_1^2) P(z_a)
\\
&&
\times 
       \frac{2x_b\xi_a\xi_b+x_a(x_b^2+\xi_b^2)}{\xi_a(\xi_b^2-x_b^2)
            (x_a\xi_b+x_b\xi_a)}
       f_g(\xi_a, Q^2)f_{\qbar'}(x_b, Q^2)
\nonumber \\
&& +\,\mbox{case: antiquarks (from $g$) come from $A$, quarks from $B$}.
\nonumber 
\end{eqnarray}

\section{NLO hard cross section}
\label{sec:NLO correction}

\subsection{Unsubtracted NLO term}
\label{sec:Unsubtrated}

The angular distribution of the lepton pair produced in Drell-Yan 
process has been calculated in \cite{Mirkes,CK} for large $Q_T$ to
order $\alpha_s$. 
The unsubtracted gluon-quark induced NLO scattering cross section corresponding
 to Fig.\ \ref{fig:gluon-quark.1} is:
\begin{eqnarray}
\label{unsubtracted.gq1}
     \frac{ d\sigma_{\rm unsubtracted \,1}(V) }
          {dQ^2 \, dy \, d\Omega \, d\xi_a \, d\xi_b }
  &=&  \frac{\alpha_s(Q^2)}{2\pi} \, \frac{s}{\hat{s}}
        \sum_{q, \, \qbar'} ~  \hat{\sigma}(V) \,
          f_{\qbar '}(\xi_b, Q^2)f_g(\xi_a, Q^2) H^{U+L}
\nonumber \\ && \hspace*{5mm}
      \times \, \frac{Q^2(Q^2+\hat{s})}{s^2(x_a\xi_b+x_b\xi_a)^2}  \Bigg \{
          (1+\cos^2 \theta) \,+ \,  \sum_{m=0}^{m=4}  
        A_m^{g\qbar}(V) G_m(\theta, \, \phi)\Bigg \}
\nonumber \\ &&
        +\,\mbox{case: antiquarks (from $g$) come from $A$, quarks from $B$},
\nonumber \\
  &=& \frac{\alpha_s(Q^2)}{2\pi}\,
        \sum_{q, \, \qbar'} ~  \hat{\sigma}(V) \,
          f_{\qbar'}(\xi_b, Q^2)f_g(\xi_a, Q^2)
\nonumber\\
  && \hspace*{5mm} \times\, 
        \left\{ -\frac{G^c(\xi_a,\, \xi_b)}{x_b - \xi_b} 
               + H^c(\xi_a, \xi_b
        \right\} 
\nonumber\\
  && \hspace*{5mm} \times\, 
        \left\{
            (1+\cos^2\theta)
            + \, \sum_{m=0}^{m=4} A_m^{g\qbar}(V) G_m(\theta, \, \phi)
        \right\}
\\ &&
      +\,\mbox{case: antiquarks (from $g$) come from $A$, quarks from $B$},
\nonumber
\end{eqnarray}
where 
\begin{eqnarray}
\label{HU} 
        H^{U+L}
        &=& \frac{(Q^2-\hat{s})^2+(Q^2-\hat{t})^2}{-\hat{s}\hat{t}} ,
\\
\label{A.02}
        A_0^{gq}(V)&=& 4 A_1^{gq}(V) =A_0^{g\qbar}(V)=4 A_1^{g\qbar}(V)
\nonumber \\ 
        &=& \frac{\hat{t}\hat{u}[(Q^2+\hat{s})^2+(Q^2-\hat{t})^2]}
                {(Q^2-\hat{t})(Q^2-\hat{u})[(Q^2-\hat{s})^2+(Q^2-\hat{t})^2]} ,
\\
\label{A.1}
        A_2^{gq}(V)&=&A_2^{g\qbar}(V)
                = \frac{1}{2\sqrt{2}}\frac{\sqrt{Q^2\hat{s}\hat{t}\hat{u}}
                [(Q^2-\hat{u})^2 - 2(Q^2-\hat{t})^2]}
                {(Q^2-\hat{t})(Q^2-\hat{u})[(Q^2-\hat{s})^2+(Q^2-\hat{t})^2]} ,
\\
\label{A.3}
        A_3^{gq}(V)&=&-A_3^{g\qbar}= A_qA_l \, \sqrt{\frac{\hat{u}\hat{t}}
        {8(Q^2-\hat{t})(Q^2-\hat{u})}} \frac{Q^4+\hat{u}^2+2\hat{s}\hat{u}}
        {(Q^2-\hat{s})^2+(Q^2-\hat{t})^2} ,
\\
\label{A.4}
        A_4^{gq}(V)
        &=&-A_4^{g\qbar}(V)
        = A_qA_l \, \sqrt{\frac{Q^2\hat{s}}{(Q^2-\hat{u})(Q^2-\hat{t})}}
        \frac{Q^4+\hat{u}^2-2Q^2\hat{t}}{(Q^2-\hat{s})^2+(Q^2-\hat{t})^2}
\nonumber \\
        &=& A_qA_l (1+\mathcal{A}^{gq})
\\
         \mathcal{A}^{gq}=\mathcal{A}^{g\qbar}
        &=& \sqrt{\frac{Q^2\hat{s}}{(Q^2-\hat{t})(Q^2-\hat{u})}}-1 
\nonumber \\ && - \,
         \sqrt{\frac{Q^2\hat{s}}{(Q^2-\hat{t})(Q^2-\hat{u})}}
         \frac{2\hat{t}(\hat{u}+\hat{t})}{(Q^2-\hat{s})^2+(Q^2-\hat{t})^2} ,
\\
\label{g0}
        G_0(\theta, \, \phi) &=& \frac{1}{2}(1-3\cos^2\theta) ,
\\
\label{g1}
        G_1(\theta, \, \phi)&=& 2 \sin^2 \theta \cos 2\phi ,
\\ \label{g2}
        G_2(\theta, \, \phi)&=& 2\sqrt{2} \sin 2\theta \cos \phi ,
\\ \label{g3}
        G_3(\theta, \, \phi)&=& 4\sqrt{2} \sin \theta \cos \phi ,
\\ \label{g4}
        G_4(\theta, \, \phi)&=& 2\cos \theta ,
\end{eqnarray}  
with $\hat{s}$, $\hat{t}$ and $\hat{u}$ being given by Eqs.\ 
(\ref{s.hat}--\ref{u.hat}). 
Also:
\begin{eqnarray}
\label{H.c}
        H^c(\xi_a,\,\xi_b) &=& 
                \frac{x_a\xi_a\xi_b^2+x_ax_b(2x_a\xi_b+x_b\xi_a)}
                {\xi_a^2\xi_b^2(x_a\xi_b+x_b\xi_a)^3}
                x_ax_b(x_ax_b+\xi_a\xi_b),
\\
\label{G.c}
        G^c(\xi_a,\,\xi_b) &=& 
        \frac{x_b(x_ax_b+\xi_a\xi_b)[(x_ax_b)^2+(x_ax_b-\xi_a\xi_b)^2]}
        {\xi_a^3\xi_b^2(x_a\xi_b+x_b\xi_a)(x_b+\xi_b)},
\end{eqnarray}
as defined in \cite{SMRS} and used in \cite{YJ}. 

Notice that, for $V=\gamma^*$, $A_q=A_l=0$,
so $A_3^{gq}(\gamma^*)=A_4^{gq}(\gamma^*)=0$, that is, only the first 4 helicity 
structure functions are nonzero for $\gamma^*$.

In the above Eqs.\ (\ref{A.02}--\ref{A.4}), only the
term containing $A_4^{Gq}(V)$ ($V=W^{\pm}, Z^0$) 
will give a collinear divergence. To get the hard cross section, we need 
to cancel out the divergence using the subtraction term obtained in Eq.\
(\ref{Collinear.gq1.new}), so we rewrite our NLO
unsubtracted parton level cross section Eq.\ (\ref{unsubtracted.gq1})
as:
\begin{eqnarray}
\label{unsubtracted.gq1.a}
     \frac{ d\sigma_{\rm unsubtracted \,1}(V) }
          {dQ^2 \, dy \, d\Omega \, d\xi_a \, d\xi_b }
  &=& \frac{\alpha_s(Q^2)}{2\pi}\,
        \sum_{q, \, \qbar'} ~  \hat{\sigma}(V) \,
          f_{\qbar'}(\xi_b, Q^2)f_g(\xi_a, Q^2)
\nonumber \\ 
&&\hspace*{5mm}
   \times\,
         \left[ -\frac{G^c(\xi_a,\, \xi_b)}{x_b - \xi_b} 
                + H^c(\xi_a, \xi_b)
         \right] 
\nonumber \\ 
&& \hspace*{5mm}
   \times\,
         \left[
              \mathcal{D}_{\qqbar'}(\theta) 
               + \, \sum_{m=0}^{m=3} A_m^{g\qbar}(V) G_m(\theta, \, \phi)
               +\, A_qA_l\mathcal{A}^{g\qbar} G_4(\theta, \, \phi)
         \right]
\nonumber\\ && \hspace*{-1cm}
    +\,\mbox{case: antiquarks (from $g$) come from $A$, quarks from $B$}.
\end{eqnarray}

\subsection{NLO term with subtraction}
\label{sec:hard}
We now subtract the showering term, Eq.\ (\ref{Collinear.gq1.new}).
Only the term proportional to $\mathcal{D}_{\qqbar'}(\theta)$ and
$\mathcal{D}_{\qbar' q} (\theta)$ needs a subtraction term; all other
terms are finite in the collinear limit ($\hat{t} \to 0 $). So we have:
\begin{eqnarray}
\label{Subtracted.gq1.New}
   \frac{ d\sigma^{\rm (New)}_{\rm hard \, 1}(V)}
        { dQ^2 \, dy \,d\Omega \, d\xi_a \, d\xi_b }
  &=& \frac{\alpha_s(Q^2)}{2\pi}\,
        \sum_{q, \, \qbar'} ~  \hat{\sigma}(V) \,
         f_g(\xi_a, Q^2)
\nonumber\\
  && \times\,\Bigg \{ \bigg [ f_{\qbar'}(\xi_b, Q^2) 
        \bigg (-\frac{G^c(\xi_a,\, \xi_b)}{x_b - \xi_b} 
        + H^c(\xi_a, \xi_b)\bigg )
\nonumber \\ && \hspace*{1cm}
         -\,f_{\qbar'}(x_b, Q^2) 
         C_1(Q_1^2) P(z_a)
       \frac{2x_b\xi_a\xi_b+x_a(x_b^2+\xi_b^2)}{\xi_a(\xi_b^2-x_b^2)
            (x_a\xi_b+x_b\xi_a)} \bigg ]
        \mathcal{D}_{\qqbar'}(\theta)
\nonumber
\\ && \hspace*{0.6cm}
     + \, f_{\qbar'}(\xi_b, Q^2)
        \bigg (-\frac{G^c(\xi_a,\, \xi_b)}{x_b - \xi_b} 
        + H^c(\xi_a, \xi_b)\bigg )
\nonumber \\ && \hspace*{12mm}
         \times \, \bigg (\sum_{m=0}^{m=3} A_m^{g\qbar}(V) G_m(\theta, \, \phi)
        +\, A_qA_l\mathcal{A}^{g\qbar} G_4(\theta, \, \phi) \bigg ) \Bigg \}
\nonumber\\ && \hspace*{-1cm}
    +\,\mbox{case: antiquarks (from $g$) come from $A$, quarks from $B$}
\\
\label{Spin}
  && \hspace*{-30mm}= \frac{\alpha_s(Q^2)}{2\pi}\,
        \sum_{q, \, \qbar'} ~  \hat{\sigma}(V) \,
         f_g(\xi_a, Q^2)f_{\qbar'}(\xi_b, Q^2) 
\nonumber\\
  && \hspace*{-25mm}\times\, 
   \Bigg (  \hat{H}^U_{g\qbar'} (1+\cos^2\theta)
\nonumber\\
  && \hspace*{-20mm}
   + \, \left\{ -\frac{G^c(\xi_a,\, \xi_b)}{x_b - \xi_b} 
                + H^c(\xi_a, \xi_b)
        \right\}
\nonumber\\
  && \hspace*{-15mm}\times\, 
    \Bigg \{ \sum_{m=0}^{m=3} A_m^{g\qbar}(V) G_m(\theta, \, \phi)
\nonumber\\
  && \hspace*{-5mm}
        +  A_qA_l \left[\mathcal{A}^{gq}+ \hat{H}^U_{g\qbar'}
           \left(-\frac{G^c(\xi_a,\, \xi_b)}
               {x_b - \xi_b} + H^c(\xi_a, \xi_b)
           \right)
           \right] G_4(\theta, \, \phi) 
   \Bigg \}
   \Bigg )
\nonumber\\ && \hspace*{-30mm}
    +\,\mbox{case: antiquarks (from $g$) come from $A$, quarks from $B$},
\end{eqnarray}
where
\begin{eqnarray}
  \hat{H}^U_{g\qbar'}&=&-\frac{G^c(\xi_a,\, \xi_b)}{x_b - \xi_b} 
        + H^c(\xi_a, \xi_b)
\nonumber \\
&& - \, \frac{f_{\qbar'}(x_b, Q^2)}{f_{\qbar'}(\xi_b, Q^2)} 
        C_1(Q_1^2) P(z_a)
        \,\frac{2x_b\xi_a\xi_b+x_a(x_b^2+\xi_b^2)}{\xi_a(x_b^2-\xi_b^2)
            (x_a\xi_b+x_b\xi_a)}.
\end{eqnarray}

Observe that, the coefficient of $\mathcal{D}_{\qqbar'}(\theta)$ agrees with 
the results in \cite{YJ}, detailed calculations shows that the pdf used here 
is the same as in \cite{YJ} and \cite{JC}.

\subsection{Results for $qg$ subprocess}
\label{sec:gq2-subprocess}

Now we specify the changes needed for the other gluon-quark scattering
subprocess, in which gluon comes out of hadron $B$ instead of $A$.  
The formulae for the subtracted cross section are obtained from Eq.\
(\ref{Subtracted.gq1.New}) by making 
the following substitutions in the right hand side of those equations: 
\begin{itemize}
\item Case that quark comes from $A$, antiquark from $B$:
\begin{equation}
         a \leftrightarrow b, ~ 1 \leftrightarrow 2, ~
        A_m^{g\qbar}(V) \leftrightarrow A_m^{qg}(V) .
\end{equation}

\item Case that antiquark comes from $A$, quark from $B$:
\begin{equation}
         a \leftrightarrow b, ~ 1 \leftrightarrow 2, ~
        A_m^{gq}(V) \leftrightarrow A_m^{\qbar g}(V) .
\end{equation}
\end{itemize}
Here 
\begin{eqnarray}
        A_m^{qg}&=&A_m^{g\qbar}( \hat{u} \leftrightarrow \hat{t})
        , ~ A_m^{\qbar g}=A_m^{gq} ( \hat{u} \leftrightarrow \hat{t})
        ,  ~(m=0,\, 1, \, 4);
\nonumber \\
        A_n^{qg}&=& -A_n^{g\qbar}( \hat{u} \leftrightarrow \hat{t})
        , ~ A_n^{\qbar g}= -A_n^{gq}( \hat{u} \leftrightarrow \hat{t})
        , ~(n=2, \,3).
\end{eqnarray}
The cut-off function $C_1(Q_1^2)$ is replaced by a function
$C_2(Q_2^2)$ with the same functional form, and we define
\begin{equation}
        -Q_2^2 = \hat{u} = s x_b \xi_b \frac{x_a^2-\xi_a^2}{x_a\xi_b+x_b\xi_a}
        , ~ z_b=\frac{x_b}{\xi_b}.
\end{equation}

\subsection{Helicity density matrix for $W^\pm$ and $Z^0$ resonance }
\label{sec:spin}
In some cases, we want to separate the contribution of different spin
configurations in the cross section, and then the helicity density
matrix is very useful.  Instead of giving the explicit formula for
helicity density matrix, we give a prescription to obtain the helicity
density matrix from the helicity cross sections.

In Eq.\ (\ref{Subtracted.gq1.New}), inside the curly brackets, let us
define $\hat{H}^U$, $\hat{H}^0$, $\hat{H}^1$, $\hat{H}^2$, $\hat{H}^3$
and $\hat{H}^P$ to be the coefficients of the angular distributions $
(1+\cos^2\theta)$, $G_0(\theta,\phi)$, $G_1(\theta,\phi)$, $G_2(\theta,\phi)$, $G_3(\theta,\phi)$
and $G_4(\theta,\phi)$.  In \cite{Mirkes}, these are expressed in terms of
what are called `helicity density matrix elements', but the
normalization condition of a density matrix, that the trace is unity,
is not satisfied.  So we will simply called them `helicity matrix
elements'.  Thus the $\hat{H}$'s are linear combination of the
helicity matrix elements.  The helicity matrix $H^{\sigma \sigma'}_{ab}$ is
then given by
\begin{eqnarray}
        H_{ab}^{00} &=& \hat{H}^0_{ab}
\nonumber \\
        H^{++}_{ab} &=& (\hat{H}_{ab}^U -  \hat{H}_{ab}^0 + 
        \hat{H}_{ab}^P ) / 2
\nonumber \\
        H^{--}_{ab} &=& (\hat{H}_{ab}^U -  \hat{H}_{ab}^0 -
        \hat{H}_{ab}^P ) / 2
\nonumber \\
        H^{+-}_{ab} &=& H^{-+}_{ab} = \hat{H}_{ab}^1
\nonumber \\
        H^{+0}_{ab} &=& H^{0+}_{ab} = \hat{H}_{ab}^3 + \hat{H}_{ab} ^2
\nonumber \\
        H^{-0}_{ab} &=& H^{0-}_{ab} = \hat{H}_{ab}^3 - \hat{H}_{ab} ^2 ,
\end{eqnarray}
where the subscript $ab$ labels the parton content of the cross
section.  Then the actual helicity density matrix is obtained from the 
above matrix by dividing it by $\hat{H}_{ab}^U$. 

In the subprocess $gq \to Vq' \to q'll'$, we have $ab = gq$, while in
the subprocess $qg \to Vq' \to q'll'$, we have $ab=qg$.

\section{Conclusion}
\label{sec:conclusion}

We have extended the application of Collins's MC algorithm to give the
angular distribution of Drell-Yan lepton pair production at order
$\alpha_s$. We also gave the corresponding helicity density matrix of $W$
and $Z$ bosons.  We confirmed that the pdf's used in Collins algorithm
are process-independent at order $\alpha_s$.

\acknowledgments

This work was supported in part by the U.S.\ Department of Energy
under grant number DE-FG02-90ER-40577.


\end{document}